\documentclass[12pt]{article}

\usepackage{pdfpages}
\usepackage[colorlinks=true,linkcolor=blue]{hyperref}
\usepackage{amsmath,amssymb,amsfonts} 
\usepackage{comment,cite}
\usepackage{color}                    
\usepackage[left=1cm,top=1cm,bottom=1cm,right=1cm,nohead,nofoot]{geometry} 

\definecolor{red}{rgb}{1,0,0}           
\definecolor{green}{rgb}{0,1,0}
\definecolor{blue}{rgb}{0,0,1}
\definecolor{Light}{gray}{.80}          
\definecolor{Dark}{gray}{.20}
\definecolor{pink}{rgb}{.95,0.82,0.92}  
\definecolor{yellow}{rgb}{1,1,0}
\definecolor{purple}{rgb}{0.85,0.15,1}
\definecolor{lightblue}{rgb}{.5,.5,1}
\definecolor{lightgray}{gray}{.87} 
\definecolor{lightyellow}{rgb}{1,1,.5}
\definecolor{darkgreen}{rgb}{0,0.5,0}
\usepackage[left=1cm,top=1cm,bottom=1cm,right=1cm,nohead,nofoot]{geometry} 

\def \be {\begin{equation}}
\def \ee {\end{equation}}
\def \bea {\begin{eqnarray}}
\def \eea {\end{eqnarray}}
\def \nn {\nonumber}

\def \rr {\raise.35ex\hbox{\small $\prime$}\kern-.17em{\mbox{\large $\imath$}}}
\def \del {\partial} 
\def \dels {\partial\kern-.6em /\kern.1em}
\def \As {{A\kern-.5em / \kern.5em}}
\def \Ds {D\kern-.7em / \kern.5em}
\def \a {\alpha}

\def \b {\beta}

\def \g {\gamma}

\def \d {\delta}
\def \eps {\epsilon}
\def \m {\mu}
\def \n {\nu}

\def \ks {k\kern-.5em /}
\def \ls {l\kern-.5em /}\def \lam {\lambda}

\def \vep {\varepsilon}
\def \II {I\hspace{-.1em}I\hspace{.1em}}

\def \IIA {\mbox{\II A\hspace{.2em}}}

\def \dd {\dot{\delta}}

\def \dm {\dot{\mu}}
\def \dn {\dot{\nu}}

\def \dlam {\dot{\lambda}}
\def \ds {\dot{\sigma}}
\def \dr {\dot{\rho}}
\def \dt {\dot{\tau}}

\def \de {\dot{\eta}}
\def \dx {\dot{\xi}}

\reversemarginpar
\newcommand{\ol}{\overline}

\begin{document}


\begin{titlepage}

\begin{center}

\hfill 
\vskip .2in

\textbf{\LARGE D-brane in R-R Field Background
}

\vskip .5in
{\large
Pei-Ming Ho$^\dagger$\footnote{
e-mail address: pmho@phys.ntu.edu.tw}, 
Chi-Hsien Yeh$^\dagger$\footnote{
e-mail address: d95222008@ntu.edu.tw}}\\
\vskip 3mm
{\it\large
$^\dagger$
Department of Physics and Center for Theoretical Sciences, \\
National Taiwan University, Taipei 10617, Taiwan,
R.O.C.}\\
\vskip 3mm
\vspace{60pt}

\end{center}
\begin{abstract}

The purpose of this paper is to understand the low energy effective theory 
of a D$p$-brane in the background of a large constant R-R $(p-1)$-form field.
We start with the M5-brane theory in large C-field background.
The $C$-field background defines a 3-dimensional volume form 
on an M5-brane,
and it is known that the low energy M5-brane theory 
can be described as a Nambu-Poisson gauge theory with 
the volume-preserving diffeomorphism symmetry (VPD).
Via a double dimensional reduction 
we obtain a D4-brane in R-R 3-form field background.
This theory has both the usual $U(1)$ gauge symmetry 
and the new symmetry of VPD.
We find that the gauge potential for VPD 
is electric-magnetic dual to the $U(1)$ gauge potential,
sharing the same physical degrees of freedom.
The result can be generalized to D$p$-branes.

\end{abstract}

\end{titlepage}

\setcounter{footnote}{0}

\section{Introduction and Motivation}

Low energy effective descriptions of D-branes and M-branes
have played a crucial role in our understanding of string theory.
They allow us to study a wide variety of subjects from AdS/CFT duality 
to brane world models.
The two basic descriptions of D-branes are 
the Dirac-Born-Infeld (DBI) theory \cite{Leigh}
and the super Yang-Mills (YM) theory \cite{Witten:1995im}.
Later it was realized that D-branes in large NS-NS $B$-field background
should be described by gauge theories on noncommutative space
\cite{CH,Schomerus,Seiberg:1999vs}.
The description of M5-branes was 
a challenging problem because of the self duality condition on gauge fields
\cite{oldM5-1,Pasti:1997gx,oldM5-2}.
A covariant DBI-like action for a single M5-brane 
was first given in \cite{Pasti:1997gx}.
More recently, 
as an analogue of noncommutative D-branes in $B$-field background,
the low energy effective theory for a single M5-brane 
in large $C$-field background was also found \cite{M51,M52}.
The latter was actually derived from 
the Bagger-Lamber-Gustavsson model \cite{BL,G}
for multiple M2-branes.
The understanding of branes often also help us understanding other branes.

The purpose of this paper is to construct new models for
D-branes in R-R field backgrounds,
to complete our understanding of D-branes in background fields.
From the viewpoint of DBI theory or the YM theory, 
one can describe an R-R field background $A^{(p+1-2n)}$
simply by adding this term
\be
\int A^{(p+1-2n)} F^n
\ee
to the D-brane action, 
where $F$ is the $U(1)$ field strength.
Why do we need to look for other descriptions?
The answer is similar to why we need noncommutative gauge theories 
for D-branes in $B$-field background.
The effect of a $B$-field can be incorporated into a D-brane action by 
simply replacing the field strength $F$ by $B+F$.
However, in the Seiberg-Witten limit \cite{Seiberg:1999vs},
the noncommutative gauge theory is a better approximation 
than the result of replacing $F$ by $B+F$ in the YM theory.
Roughly speaking, when the $B$-field is large enough, 
higher derivative terms that would normally 
be ignored in a low energy effective theory
can no longer be ignored if it is multiplied by a certain power of $B$.
We would like to understand analogous effects of R-R fields on D-branes.

This problem has been studied in various aspects 
via different approaches.
In \cite{Cornalba:2002cu}, it was shown how R-R background potential
influences D-brane dynamics in a way consistent with S-duality, 
so that Moyal deformation can be induced by R-R potential as well 
as the NS-NS $B$-field background.
In \cite{Ooguri:2003qp}, 
it was shown that the anti-commutation relation of fermionic fields 
can be modified by a graviphoton background, 
and the result was generalized to generic R-R backgrounds 
in \cite{deBoer:2003dn}.
In this paper we take yet another approach and find that 
a generalized Nambu-Poisson structure is induced by the R-R background, 
characterizing a new gauge symmetry -- 
the volume-preserving diffeomorphism -- on the D-brane.

The noncommutativity on a D-brane due to a $B$-field background 
can roughly be understood as the effect due to the coupling 
of the $B$-field to an open string ending on the D-brane.
Similarly, an R-R $(k+1)$-form gauge potential couples to a D$k$-brane
ending on a D$p$-brane,
and interaction of excitations on a D$p$-brane mediated by a D$k$-brane 
would be influenced by the R-R field background.
Sufficiently strong R-R backgrounds can thus turn on new interactions 
usually ignored in a low energy effective theory.

Instead of computing directly the dynamics of D-branes ending on D-branes, 
our strategy is to fully utilize string dualities.
The starting point is the above-mentioned new M5-brane theory \cite{M51,M52} 
in a large $C$-field background.
The $C$-field background defines a 3-dimensional volume form,
and the M5-brane theory is a gauge theory of diffeomorphisms 
preserving this volume form.
We will refer to this theory as the Nambu-Poisson (NP) M5-brane theory
for the gauge algebra is defined through the Nambu-Poisson bracket \cite{Nambu}. 
Various calculations \cite{Ho:2007vk} suggest
that Nambu-Poisson bracket appears in a $C$-field background 
for open membranes in the same fashion that 
Moyal bracket appears in a $B$-field background for open strings.

An M5-brane is related to a D4-brane through double dimensional reduction (DDR), 
which means the simultaneous compactification of a direction 
in the target space and a direction on an M5-brane.
The $C$-field in M theory leads to either a 3-form R-R gauge potential 
and/or a 2-form NS-NS $B$-field in the type \IIA theory 
after compactification,
depending on the direction of the compactified circle.
In \cite{M52}, the compactified circle is chosen such that 
the $C$-field background reduces to a $B$-field background, 
and the NP M5-brane theory reduces to 
the Poisson limit of the noncommutative gauge theory of a D4-brane.
\footnote{
The Poisson limit of a noncommutative algebra refers to 
the approximation of the Moyal product 
by the leading order correction to the commutative product.
In this limit the commutator of Moyal product 
reduces to the Poisson bracket.
}
In particular,
the Nambu-Poisson bracket in the NP M5-brane theory
reduces to the Poisson bracket.
This can be viewed as an evidence for the validity of 
the NP M5-brane theory. 
Another evidence was obtained by examining 
the self dual string solutions corresponding to an M2-brane
ending on an M5-brane \cite{Furuuchi:2009zx}.
A short review of the NP M5-brane was given in \cite{HoM5}.

In this article we will carry out DDR in another direction 
so that the $C$-field background is reinterpreted as 
a constant R-R 3-form gauge potential.
We will use the same symbol $C$ to refer to both 
the M theory 3-form gauge potential and 
the 3-form R-R potential in type \IIA string theory.
The first goal of this paper is to understand
the D4-brane theory in a constant $C$-field background.
It is expected that the geometry of this theory is equipped 
with a 3-bracket structure \cite{ChuSmith,Huddleston:2010cx}.

The DDR of the NP M5-brane to D4-brane is highly nontrivial.
The gauge symmetry of an NP M5-brane is 
the volume-preserving diffeomorphisms (VPD).
Since the $C$-field background is parallel 
to the D4-brane,
it is natural to expect that the D4-brane inherits 
the VPD symmetry.
However, the massless spectrum of a D4-brane 
(a $U(1)$ gauge potential $A$, 5 scalars $\phi$ and 
their fermionic superpartners) 
does not include a 2-form gauge potential for the VPD gauge symmetry.
We will show in the following that, interestingly,
the 2-form gauge potential 
is dual to the 1-form $U(1)$ potential $A$, 
sharing the same physical degrees of freedom. 
While the VPD algebra is non-Abelian, 
the mathematical description of this duality is not straightforward at all.

The electric-magnetic duality between $U(1)$ gauge theory and VPD 
can be understood physically as follows. 
The endpoint of a fundamental string is an electric charge on a D4-brane,
and the massless fluctuation of an open string in the longitudinal directions 
of the D4-brane constitutes the $U(1)$ gauge potential.
From the M5-brane's viewpoint,
the VPD gauge potential comes from the massless excitations of M2-branes 
ending on an M5-brane. 
It is a 2-form potential because the boundary of an M2-brane is a string.
Therefore, from the D4-brane's viewpoint, 
the VPD gauge potential is associated with the boundary of a D2-brane,
which is interpreted as the magnetic charge on the D4-brane, 
and so we expect the electric-magnetic duality
between the $U(1)$ symmetry and VPD.

The interpretation above can be easily generalized to other D$p$-branes.
The endpoint of a fundamental string is an electric charge on the D$p$-brane.
The magnetic charge is then the boundary of a D(p-2)-brane 
ending on the D$p$-brane.
Massless fluctuations of the D$(p-2)$-brane in the longitudinal directions 
give rise to the $(p-2)$-form potential for the VPD
of the $(p-1)$ dimensional volume 
defined by an R-R $(p-1)$-form background.
The corresponding $(p-1)$-form field strength is then dual to
the 2-form field strength of the $U(1)$ symmetry.
The second goal of this paper is to construct gauge theories
describing a single D$p$-brane in constant R-R $(p-1)$-form field background.
The results give us hints about D$p$-brane theories in
more general backgrounds of R-R fields.
We leave this topic for future study.

The plan of this paper is as follows.
We give a brief review of the NP M5-brane theory in Sec. \ref{review}.
In Sec. \ref{DDR} we derive the D4-brane action 
in large $C$-field background from the NP M5-brane action 
via double dimensional reduction, 
and in Sec. \ref{CovVar} we define 2-form field strengths $F$
such that they are not only invariant under the $U(1)$ gauge transformations
but also covariant under the VPD.
We study the 0-th order and 1st order terms of the D4-brane action 
in the perturbative expansion in Sec. \ref{ExpandD4} 
to show how the new action differs from Maxwell's action.
For simplicity, matter fields are ignored in the calculation except in Sec. \ref{Matter}.
We generalize the gauge field theory to multiple D$p$-branes for generic $p$ 
in Sec. \ref{Generalization}. 
Finally we conclude in Sec. \ref{Conclusion}.

\section{Review of Nambu-Poisson M5-brane Theory}
\label{review}

The worldvolume theory of the M5-brane has the field content 
of a self-dual 2-form gauge potential ($b_{\dm\dn}, b_{\dm\nu}$), 
5 scalars ($X^i$) and the dimensional reduction of 
an 11 dimensional Majorana fermion ($\Psi$).
\footnote{
The fermion $\Psi$ is chiral, i.e., $\Gamma^7 \Psi = -\Psi$, 
where $\Gamma^7$ is defined by
$\Gamma^7 \equiv
\Gamma^{012}\Gamma^{\dot{1}\dot{2}\dot{3}}$.
}
The world volume coordinates will be denoted as
$\{ x^{\mu}, y^{\dm} \} = \{ x^0, x^1, x^2, y^{\dot{1}}, y^{\dot{2}}, y^{\dot{3}} \}$.
In a $C$-field background
the M5-brane action should respect 
the worldvolume translational symmetry, 
the global $SO(2,1)\times SO(3)$ rotation symmetry, 
the gauge symmetry of volume-preserving diffeomorphisms 
and the 6D ${\cal N}$ = (2, 0) supersymmetry.
In the limit $\eps\rightarrow 0$ \cite{Chen:2010ny} with
\bea
&\ell_P \sim \eps^{1/3}, \qquad
g_{\mu\nu} \sim 1, \qquad
g_{\dot{\mu}\dot{\nu}} \sim \eps, \qquad
C_{\dm\dn\dlam} \sim 1, 
\label{NPlimit} \\
&(\mu, \nu = 0, 1, 2 \;\; \mbox{and} \;\; 
\dm, \dn, \dlam = \dot{1}, \dot{2}, \dot{3})
\nn
\eea
a good approximation of the M5-brane in $C$-field background
is given by the action \cite{M52}
\footnote{
This action was first derived from the Bagger-Lambert action \cite{BL}
with the Lie 3-algebra chosen to be the Nambu-Poisson algebra \cite{M51,M52}.
}
\be
S = 
S_{X} + S_{\Psi} + S_{gauge},
\qquad
S_{gauge} = S_{{\cal H}^2} 
+ S_{CS},
\label{M5S}
\ee
where
\footnote{
$\Psi$ here was denoted by $\Psi'$ in \cite{M52}.
}
\begin{eqnarray}
S_{X}
&=&\int d^3 x d^3 y \; \left[
-\frac{1}{2}({\cal D}_\mu X^i)^2
-\frac{1}{2}({\cal D}_{\dot\lambda}X^i)^2
\right.\nonumber\\&&\left.
-\frac{1}{2g^2}
-\frac{g^4}{4}\{X^{\dot\mu},X^i,X^j\}^2
-\frac{g^4}{12}\{X^i,X^j,X^k\}^2\right], 
\label{Sboson}
\\
S_{\Psi}
&=&\int d^3 x d^3 y \; \left[
\frac{i}{2}\ol\Psi\Gamma^\mu {\cal D}_\mu\Psi
+\frac{i}{2}\ol\Psi\Gamma^{\dot\rho}{\cal D}_{\dot\rho}\Psi
\right.\nonumber\\&&\left.
+\frac{ig^2}{2}\ol\Psi\Gamma_{\dot\mu i}\{X^{\dot\mu},X^i,\Psi\}
-\frac{ig^2}{4}\ol\Psi\Gamma_{ij}\Gamma_{\dot1\dot2\dot3}\{X^i,X^j,\Psi\}
\right], 
\label{Sfermi0} 
\\
S_{{\cal H}^2} 
&=&\int d^3 x d^3 y \; \left[
-\frac{1}{12}{\cal H}_{\dot\mu\dot\nu\dot\rho}^2
-\frac{1}{4}{\cal H}_{\lambda\dot\mu\dot\nu}^2
\right],
\\
S_{CS}
&=&
\int d^3 x d^3 y \; 
\epsilon^{\mu\nu\lambda}\epsilon^{\dot\mu\dot\nu\dot\lambda}
\left[ -\frac{1}{2}
\partial_{\dot\mu}b_{\mu\dot\nu}\partial_\nu b_{\lambda\dot\lambda}
+\frac{g}{6}
\partial_{\dot\mu}b_{\nu\dot\nu}
\epsilon^{\dot\rho\dot\sigma\dot\tau}
\partial_{\dot\sigma}b_{\lambda\dot\rho}
(\partial_{\dot\lambda}b_{\mu\dot\tau}-\partial_{\dot\tau}b_{\mu\dot\lambda})
\right].
\label{CSt}
\end{eqnarray}
In the above we use the notation
\bea
X^{\dot\mu}(y)
&\equiv&\frac{y^{\dot\mu}}{g}+\frac{1}{2}\epsilon^{\dot\mu\dot\kappa\dot\lambda}
b_{\dot\kappa\dot\lambda}(y),\\
\{A,B,C\}&\equiv&\eps^{\dm\dn\dr}\del_{\dm}A\del_{\dn}B\del_{\dr}C.
\eea
The overall coefficient of the action $S$ has been scaled to $1$
by rescaling the fields and worldvolume coordinates.

For the matter fields $X^i, \Psi$,
the covariant derivatives are defined by 
\footnote{
Here and below we use $\Phi$ to represent
both matter fields $X^i, \Psi$.
}
\bea
{\cal D}_\mu\Phi
&\equiv&\partial_\mu\Phi
-g\{b_{\mu\dot\nu},y^{\dot\nu},\Phi\},
\qquad
\label{dmu}
\\
{\cal D}_{\dot\mu}\Phi
&\equiv&\frac{g^2}{2}\epsilon_{\dot\mu\dot\nu\dot\rho}
\{X^{\dot\nu},X^{\dot\rho},\Phi\}. 
\label{ddotmu} 
\eea

The definition of the 3-form field strength as
\begin{eqnarray}
H_{\lambda\dot\mu\dot\nu}
&=&
\partial_{\lambda}b_{\dot\mu\dot\nu}
-\partial_{\dot\mu}b_{\lambda\dot\nu}
+\partial_{\dot\nu}b_{\lambda\dot\mu},\\
H_{\dot\lambda\dot\mu\dot\nu}
&=&
\partial_{\dot\lambda}b_{\dot\mu\dot\nu}
+\partial_{\dot\mu}b_{\dot\nu\dot\lambda}
+\partial_{\dot\nu}b_{\dot\lambda\dot\mu}
\end{eqnarray}
is no longer covariant under the non-Abelian gauge transformations.
The covariant 3-form field strengths ${\cal H}$ 
should be defined as
\begin{eqnarray}
{\cal H}_{\lambda\dot\mu\dot\nu}
&=&\epsilon_{\dot\mu\dot\nu\dot\lambda}{\cal D}_\lambda X^{\dot\lambda}
\nonumber\\
&=&H_{\lambda\dot\mu\dot\nu}
-g\epsilon^{\dot\sigma\dot\tau\dot\rho}
(\partial_{\dot\sigma}b_{\lambda\dot\tau})
\partial_{\dot\rho}b_{\dot\mu\dot\nu},\label{h12def}\\
{\cal H}_{\dot1\dot2\dot3}
&=&g^2\{X^{\dot1},X^{\dot2},X^{\dot3}\}-\frac{1}{g}
\nonumber\\
&=&H_{\dot1\dot2\dot3}
+\frac{g}{2}
(\partial_{\dot\mu}b^{\dot\mu}\partial_{\dot\nu}b^{\dot\nu}
-\partial_{\dot\mu}b^{\dot\nu}\partial_{\dot\nu}b^{\dot\mu})
+g^2\{b^{\dot1},b^{\dot2},b^{\dot3}\}.
\label{h30def}
\end{eqnarray}

The fundamental fields transform under the gauge transformation as
\begin{eqnarray}
\delta_{\Lambda}\Phi
&=&g \kappa^{\dot\rho}\partial_{\dot\rho}\Phi \label{gt1}
\qquad
(\Phi = X^i, \Psi),
\\
\delta_{\Lambda}b_{\dot\kappa\dot\lambda}
&=&\partial_{\dot\kappa}\Lambda_{\dot\lambda}
-\partial_{\dot\lambda}\Lambda_{\dot\kappa}
+g\kappa^{\dot\rho}\partial_{\dot\rho} b_{\dot\kappa\dot\lambda},
\label{gt2}
\\
\delta_{\Lambda} b_{\lambda\dot\sigma}
&=&\partial_\lambda\Lambda_{\dot\sigma}
-\partial_{\dot\sigma}\Lambda_\lambda
+g\kappa^{\dot\tau}\partial_{\dot\tau}b_{\lambda\dot\sigma}
+g(\partial_{\dot\sigma}\kappa^{\dot\tau})b_{\lambda\dot\tau}, \label{gt4}
\end{eqnarray}
where
\be
\kappa^{\dot\lambda}\equiv
\epsilon^{\dot\lambda\dot\mu\dot\nu}\partial_{\dot\mu}
\Lambda_{\dot\nu}(x,y).
\ee
The field strengths ${\cal H}$ transform like $\Phi$.

The gauge transformations
can be more concisely expressed in terms of 
the new variables $b^{\dm}, B_{\mu}{}^{\dm}$
\bea
b^{\dm} &\equiv& \frac{1}{2}\eps^{\dm\dn\dlam} b_{\dn\dlam}, \\
B_{\mu}{}^{\dm} &\equiv& \eps^{\dm\dn\dlam}\del_{\dn}b_{\mu\dlam}
\label{B}
\eea
for the gauge fields as 
\bea
\d_{\Lambda} b^{\dm} &=& \kappa^{\dm} + g\kappa^{\dn}\del_{\dn} b^{\dm}, 
\\
\d_{\Lambda} B_{\mu}{}^{\dm} &=& 
\del_{\mu}\kappa^{\dm} + g\kappa^{\dn}\del_{\dn}B_{\mu}{}^{\dm}
- g(\del_{\dn}\kappa^{\dm})B_{\mu}{}^{\dn}.
\eea
In terms of $B_{\mu}{}^{\dn}$, 
the covariant derivative ${\cal D}_{\mu}$ acts as 
\be
{\cal D}_{\mu} \Phi = \del_{\mu} \Phi - gB_{\mu}{}^{\dm}\del_{\dm} \Phi.
\ee
Remarkably, the field $b_{\mu\dm}$ appears in the action 
only through the variable $B_{\mu}{}^{\dm}$.

Another feature of the gauge transformations
is that, in terms of $X^i, \Psi, b^{\dm}$ and $B_{\mu}{}^{\dm}$, 
all gauge transformations can be expressed solely in terms of $\kappa^{\dm}$, 
without referring to $\Lambda_{\dm}$,
as long as one keeps in mind the constraint 
\be
\del_{\dm}\kappa^{\dm} = 0.
\ee
This gauge transformation can be naturally interpreted as 
volume-preserving diffeomorphism (VPD)
\be
\d y^{\dm} = g\kappa^{\dm}, 
\qquad \mbox{with} \qquad
\del_{\dm} \kappa^{\dm} = 0.
\ee
The field $b^{\dm}$ is then interpreted as 
the gauge potential for the VPD 
in the 3 dimensional space picked by the $C$-field background.

The M5-brane theory is also invariant under
the supersymmetry transformations $\d^{(1)}_{\chi}$, $\d^{(2)}_{\eps}$.
We have
\begin{equation}
\delta^{(1)}_\chi \Psi=\chi,\quad
\delta^{(1)}_\chi X^i=\delta^{(1)}_\chi b_{\dot\mu\dot\nu}
=\delta^{(1)}_\chi b_{\mu\dot\nu}=0,
\label{chitr}
\end{equation}
and
\footnote{
$\eps$ here was denoted by $\eps'$ 
in \cite{M52}.
}
\begin{eqnarray}
\delta^{(2)}_\eps X^i
&=&i\ol\epsilon\Gamma^i\Psi,
\label{dX}
\\
\delta^{(2)}_\eps \Psi
&=&{\cal D}_\mu X^i\Gamma^\mu\Gamma^i\epsilon
+{\cal D}_{\dot\mu}X^i\Gamma^{\dot\mu}\Gamma^i\epsilon
\nonumber\\&&
-\frac{1}{2}
{\cal H}_{\mu\dot\nu\dot\rho}
\Gamma^\mu\Gamma^{\dot\nu\dot\rho}\epsilon
-\frac{1}{g}\left(1+g{\cal H}_{\dot1\dot2\dot3}\right)
\Gamma_{\dot1\dot2\dot3}\epsilon
\nonumber\\&&
-\frac{g^2}{2}\{X^{\dot\mu},X^i,X^j\}
\Gamma^{\dot\mu}\Gamma^{ij}\epsilon
+\frac{g^2}{6}\{X^i,X^j,X^k\}
\Gamma^{ijk}\Gamma^{\dot1\dot2\dot3}\epsilon,
\label{dPsi}
\\
\delta^{(2)}_\eps b_{\dot\mu\dot\nu}
&=&-i(\ol\epsilon\Gamma_{\dot\mu\dot\nu}\Psi),
\label{db1}
\\
\delta^{(2)}_\eps b_{\mu\dot\nu}
&=&-i\left(1+g{\cal H}_{\dot1\dot2\dot3}\right)
\ol\epsilon\Gamma_\mu\Gamma_{\dot\nu}\Psi
+ig(\ol\epsilon\Gamma_\mu\Gamma^i\Gamma_{\dot1\dot2\dot3}\Psi)
\partial_{\dot\nu}X^i.
\label{db2}
\end{eqnarray}
The SUSY transformation parameters $\chi$, $\eps$ can be 
conveniently denoted as an 11D Majorana spinor
satisfying the 6D chirality condition
\be
\Gamma^7\chi=\chi, \qquad
\Gamma^7\epsilon=\epsilon.
\ee
They are both nonlinear SUSY transformations,
but a superposition of the two,
\be
\d^{(1)}_\chi + g \d^{(2)}_\eps 
\qquad \mbox{with} \qquad
\chi = \Gamma^{\dot{1}\dot{2}\dot{3}}\eps,
\ee
defines a linear SUSY transformation.

\section{D4-Brane via Double Dimensional Reduction}
\label{DDR}

To carry out the double dimensional reduction (DDR) 
for the M5-brane along the $x^2$-direction,
we set 
\be
x^2 \sim x^2 + 2\pi R,
\ee
and let all other fields to be independent of $x^2$.
As a result we can set $\del_2$ to zero when it acts on any field.
Here $R$ is the radius of the circle of compactification 
and we should take $R \ll 1$
such that the 6 dimensional field theory on M5 
reduces to a 5 dimensional field theory for D4.
Since the NP M5-brane action (\ref{M5S}) is a good low energy 
effective theory in the limit (\ref{NPlimit}),
the 5 dimensional field theory is a good low energy effective 
description of a D4-brane in the limit $\eps\rightarrow 0$ for
\be
\ell_s \sim \eps^{1/2}, \qquad
g_s \sim \eps^{-1/2}, \qquad
g_{\a\b} \sim 1, \qquad
g_{\dm\dn} \sim \eps, \qquad
C_{\dm\dn\dlam} \sim 1,
\label{D4limit}
\ee
with 
\be
g_s \ell_s \ll 1,
\label{gl}
\ee
from the perspective of the type \IIA theory.
The indices $\a,\b = 0,1$ are used to 
distinguish from the M5-brane indices $\m,\n = 0, 1, 2$.

Note that in the limit (\ref{NPlimit})
the $C$-field component $C_{012} \sim \eps^{-1}$.
As a result the $B$-field component $B_{01} \sim \eps^{-1}$
and the noncommutative parameter $\theta^{01} \sim B^{-1} \sim \eps$ 
vanishes in the limit $\eps \rightarrow 0$.
However, the combination $2\pi\a' B$ is finite in the limit, 
and thus the D4-brane is not only in a $C$-field background 
but also in the $B$-field background.
Using the nonlinear self-dual relation derived in \cite{oldM5-1},
we can express $C_{012}$ in terms of $C_{\dot{1}\dot{2}\dot{3}}$,
and then the $B$-field background is given by
\be
2\pi \a' B_{01} 
= \frac{C_{\dot{1}\dot{2}\dot{3}}}{2\pi}.
\ee
In the convention (normalization of the worldvolume coordinates) 
of \cite{M52},
we have
\be
C_{\dot{1}\dot{2}\dot{3}} = \frac{1}{g^2} \qquad
\Rightarrow \qquad
2\pi \a' B_{01} = \frac{1}{2\pi g^2}.
\label{B-background}
\ee

\subsection{D4-brane Action for the Gauge Fields}

For simplicity let us ignore the matter fields 
for the time being, 
and focus on the gauge field part of the action $S_{gauge}$.
We will give the full action including matter fields later in Sec. \ref{Matter}.
The result of DDR on $S_{gauge}$ is
\bea
S^{(1)}_{gauge}[b^{\dm},a_{\dm},b_{\a\dm}]&=&\int d^{2}x d^{3}y
\left\{ -\frac{1}{2}{\cal H}_{\dot{1}\dot{2}\dot{3}}^{2} 
-\frac{1}{4}{\cal H}_{2\dm\dn}^{2} 
-\frac{1}{4}{\cal H}_{\a\dm\dn}^2 \right. \nn\\
&&+\eps^{\a\b}\eps^{\dm\dn\dr}\del_{\b}a_{\dr}\del_{\dm}b_{\a\dn} 
\left.+\frac{g}{2}\eps^{\a\b}\eps_{\dm\dn\dr}\eps^{\dm\dd\dt}
\eps^{\dn\ds\dlam}\eps^{\dr\de\dx}
\del_{\dd}b_{\a\dt}\del_{\ds}b_{\b\dlam}\del_{\de}a_{\dx}\right\},
\label{Sp}
\eea
where
\be 
a_{\dm} \equiv b_{\dm 2}.
\ee
Apparently we should identify $a_{\dm}$ as components 
of the one-form potential on the D4-brane.
In terms of the field strength
\be
F_{\dm\dn} \equiv \del_{\dm}a_{\dn}-\del_{\dn}a_{\dm},
\ee
we can rewrite ${\cal H}_{2\dm\dn}$ as
\be
{\cal H}_{2\dm\dn} = F_{\dm\dn} + 
\frac{g}{2}\eps_{\dm\dn\dlam}\eps^{\ds\dr\dt}\del_{\ds}b^{\dlam}F_{\dr\dt}.
\label{H2dmdn}
\ee

In the above we see that part of the two-form potential $b$
on the M5-brane transforms into part of the one-form potential $a$ on D4.
However, in order to interpret this action as a D4-brane action, 
we still need to identify the rest of the components $a_{\a}$
of the one-form gauge potential, 
and to re-interpret $b_{\a\dm}$ and $b_{\dm\dn}$
from the D4-brane viewpoint.
We expect that the $U(1)$ gauge symmetry on the D4-brane 
has its origin in the gauge transformations (\ref{gt2}), (\ref{gt4}) 
on the M5-brane.
The gauge transformation parameter $\Lambda_2$
shall be identified with the $U(1)$ gauge transformation parameter.
This is consistent with the identification of $a_{\dm}$ with $b_{\dm 2}$.
The gauge symmetry parametrized by $\Lambda_{\dm}$, i.e.,
the VPD, is also still present on the D4-brane.

\subsection{Duality Transformation}

In order to understand the physical meaning of the action (\ref{Sp}), 
we try to simplify the action 
by integrating out the remaining components of 
the 2-form gauge field $b$ as much as possible, 
since there is no 2-form gauge potential 
in the usual description of a D4-brane.

First we note that the action (\ref{Sp}) depends on 
$b_{\a\dm}$ only through the variable $B_{\a}{}^{\dm}$ (\ref{B}).
In terms of $B_{\a}{}^{\dm}$, we have 
\be
{\cal H}_{\a\dm\dn} =
\eps_{\dm\dn\dlam}(\del_{\a}b^{\dlam}-V_{\ds}{}^{\dlam}B_{\a}{}^{\ds}),
\label{Hadmdn}
\ee
where
\be
V_{\dn}^{~\dm}\equiv\d_{\dn}^{~\dm}+g\del_{\dn}b^{\dm}.
\label{def-V}
\ee
Hence we can rewrite the action (\ref{Sp}) as
\bea
S^{(2)}[b^{\dm},a_{\dm},B_{\a}^{~\dm}]&=&
\int d^{2}x d^{3}y \left\{ 
-\frac{1}{2}{\cal H}_{\dot{1}\dot{2}\dot{3}}^{2} 
-\frac{1}{4}{\cal H}_{2\dm\dn}^2
\right. \nn\\
&&-\frac{1}{2}(\del_{\a}b^{\dm}-V_{\ds}^{~\dm}B_{\a}^{~\ds})^2
\left. +\eps^{\a\b}\del_{\b}a_{\dm}B_{\a}^{~\dm}
+\frac{g}{2}\eps^{\a\b}F_{\dm\dn}B_{\a}^{~\dm}B_{\b}^{~\dn}\right\}.
\label{Spp}
\eea

It turns out that it is possible to extract the components $a_{\a}$
on the D4-brane by dualizing the field $B_{\a}^{~\dm}$.
We can introduce the Lagrange multiplier $f_{\a\dm}$ 
to rewrite the action (\ref{Spp}) as
\bea
S^{(3)}[b^{\dm},a_{\dm},b_{\a\dm},\breve{B}_{\a}^{~\dm},f_{\b\dm}]&=&
\int d^{2}x d^{3}y \left\{ -\frac{1}{2}{\cal H}_{\dot{1}\dot{2}\dot{3}}^{2} 
-\frac{1}{4}{\cal H}_{2\dm\dn}^2
-\frac{1}{2}(\del_{\a}b^{\dm}-V_{\ds}^{~\dm}\breve{B}_{\a}^{~\ds})^2
\right. 
\nn\\
&&
\left. +\eps^{\a\b}\del_{\b}a_{\dm}\breve{B}_{\a}^{~\dm}
+\frac{g}{2}\eps^{\a\b}F_{\dm\dn}\breve{B}_{\a}^{~\dm}\breve{B}_{\b}^{~\dn}
-\eps^{\a\b}f_{\b\dm}
[\breve{B}_{\a}^{~\dm}-\eps^{\dm\dn\dr}\del_{\dn}b_{\a\dr}]\right\}, 
\nn \\
\eea
where we used the notation $\breve{B}$ for a new variable 
independent of $b_{\a\dm}$.
If we integrate out the Lagrange multiplier $f_{\b\dm}$, 
we will get $\breve{B}_{\a}^{~\dm}=B_{\a}^{~\dm}$, 
and the action above reduces back to (\ref{Spp}).

Instead, we can integrate out $\breve{B}_{\a}^{~\dm}$ and $b_{\a\dm}$ 
to dualize the field $B_{\a}^{~\dm}$.
First we integrate out $b_{\a\dm}$, 
and find the constraint on $f_{\a\dm}$
\be
\eps^{\dm\dn\dlam}\del_{\dm}f_{\a\dn}=0.
\ee
It implies that, locally
\be
f_{\a\dm}=\del_{\dm}a_{\a}
\ee
for some potential $a_{\a}$.
Hence, after integrating out $b_{\a\dm}$, we get
\bea
S^{(4)}[b^{\dm},a_{\dm},a_{\a},\breve{B}_{\a}^{~\dm}]&=&
\int d^{2}x d^{3}y \left\{ 
-\frac{1}{2}{\cal H}_{\dot{1}\dot{2}\dot{3}}^{2} 
-\frac{1}{4}{\cal H}_{2\dm\dn}^2
-\frac{1}{2}(\del_{\a}b^{\dm}-V_{\ds}^{~\dm}\breve{B}_{\a}^{~\ds})^2
\right.
\nn\\
&&+\eps^{\a\b}\del_{\b}a_{\dm}\breve{B}_{\a}^{~\dm}
+\frac{g}{2}\eps^{\a\b}F_{\dm\dn}\breve{B}_{\a}^{~\dm}\breve{B}_{\b}^{~\dn}
\left.-\eps^{\a\b}\del_{\dm}a_{\b}\breve{B}_{\a}^{~\dm}\right\}.
\label{S4}
\eea

The next step is to integrate out $\breve{B}_{\a}^{~\dm}$ 
to get the dual action.
Since the action is at most quadratic in $\breve{B}_{\a}^{~\dm}$,
the result of integrating out $\breve{B}_{\a}^{~\dm}$
is the same as replacing $\breve{B}_{\a}^{~\dm}$ by the solution to
its equation of motion, which is a constraint
\be
V_{\dm}^{~\dn}(\del^{\a}b_{\dn}
-V^{\dr}_{~\dn}\breve{B}^{\a}_{~\dr})
+\eps^{\a\b}F_{\b\dm}
+g\eps^{\a\b}F_{\dm\dn}\breve{B}_{\b}^{~\dn}=0.
\ee
The solution of $\breve{B}_{\a}^{~\dm}$, denoted as 
$\hat{B}_{\a}^{~\dm}$, is given by
\be
\hat{B}_{\a}^{~\dm} \equiv
(M^{-1})_{\a\b}{}^{\dm\dn}
(V_{\dn}^{~\ds}\del^{\b}b_{\ds}+\eps^{\b\g}F_{\g\dn}),
\label{def-B}
\ee
where 
\be
M_{\dm\dn}{}^{\a\b} \equiv
V_{\dm\dr}V_{\dn}{}^{\dr}\d^{\a\b}-g\eps^{\a\b}F_{\dm\dn},
\label{def-M}
\ee
and $M^{-1}$ is defined by
\be
(M^{-1})_{\g\a}{}^{\dlam\dm}M_{\dm\dn}{}^{\a\b}=\d^{\dlam}{}_{\dn}\d^{~\b}_{\g}.
\ee

After integrating out $\breve{B}_{\a}^{~\dm}$, we get
\bea
S^{(5)}[b^{\dm},a_{\dm},a_{\a}]&=&
\int d^{2}x d^{3}y \left\{ 
-\frac{1}{2}{\cal H}_{\dot{1}\dot{2}\dot{3}}^{2}
-\frac{1}{4}(F_{\dn\dr}+
\frac{g}{2}\eps_{\dm\dn\dr}\eps^{\ds\dd\dt}\del_{\ds}b^{\dm}F_{\dd\dt})^{2}
-\frac{1}{2}\del_{\a}b^{\dm}\del^{\a}b_{\dm}
\right. \nn\\
&&\left.
+\frac{1}{2}(\eps^{\a\g}F_{\g\dm}+V_{\dm}^{~\ds}\del^{\a}b_{\ds})
(M^{-1})_{\a\b}{}^{\dm\dn}(\eps^{\b\d}F_{\d\dn}+
V_{\dn}{}^{\dlam}\del^{\b}b_{\dlam}) \right\}.
\label{dual action}
\eea 

At the quantum level, there is a one-loop contribution to 
the action when we integrate out $\breve{B}_{\a}{}^{\dm}$.
It is
\be
\Delta S_{1-loop} = -\frac{\hbar}{2}Tr(Log(M_{\dm\dn}{}^{\a\b})).
\ee

The action (\ref{dual action}) is only remotely resembling
the familiar Maxwell action 
for a $U(1)$ gauge theory we expect on the D4-brane.
We can find terms resembling $F_{\dm\dn}^2$ and $F_{\a\dm}^2$,
but the coefficients do not match.
The term $F_{\a\b}^2$ is missing.
We still have the field $b^{\dm}$ which can not be easily 
integrated out because it has 2nd derivative terms in the action.
It appears that we need to keep the field $b^{\dm}$, 
which continues to play the role of the gauge potential 
for the gauge transformation parametrized by $\Lambda_{\dm}$,
but we need to identify its physical degrees of freedom 
in the D4-brane theory.

Having decided to keep the gauge transformations 
parametrized by $\Lambda_{\dm}$ as a new gauge symmetry 
in the D4-brane theory,
we need to define covariant field strengths
suitable for the gauge transformations.

\section{Covariant Variables}
\label{CovVar}

\subsection{Gauge Transformation}

The gauge transformations of $b^{\dm}$ and $a_{\dm} = b_{\dm 2}$ 
are inherited from the NP M5-brane theory as
\bea
\delta_{\Lambda} b^{\dm}&=&
\kappa^{\dm}+g\kappa^{\dn}\del_{\dn}b^{\dm},
\label{transf-bdm}\\
\delta_{\Lambda} a_{\dm}&=&
\del_{\dm}\lambda+
g(\kappa^{\dn}\del_{\dn}a_{\dm}+a_{\dn}\del_{\dm}\kappa^{\dn}),
\label{transf-adm}
\eea
where $\lam \equiv \Lambda_2$.

The field $a_{\a}$ was introduced by hand and
so its gauge transformation rule has to be solved from 
the requirement that the action $S^{(4)}$ (\ref{S4}) be invariant. 
For a quick derivation one needs to realize that the Chern-Simons term 
must be gauge invariant by itself.
Plugging in the gauge transformation of $\breve{B}_{\a}^{~\dm}$
\footnote{
The gauge transformation of $\breve{B}_{\a}^{~\dm}$
should be the same as that of $B_{\a}^{~\dm}$.
}
and $b^{\dm}$,
the gauge transformation of the CS term (after integration by part ) is
\bea
&&\delta_{\Lambda}(\eps^{\a\b}\del_{\b}a_{\dm}\breve{B}_{\a}^{~\dm}
+\frac{g}{2}\eps^{\a\b}F_{\dm\dn}\breve{B}_{\a}^{~\dm}\breve{B}_{\b}^{~\dn}
-\eps^{\a\b}\del_{\dm}a_{\b}\breve{B}_{\a}^{~\dm})\nn\\
&=&\del_{\dm}\breve{B}_{\a}^{~\dm}\eps^{\a\b}[-\del_{\b}\lam
-g(\kappa^{\ds}\del_{\ds}a_{\b}+a_{\ds}\del_{\b}\kappa^{\ds}) 
+ \d a_\b].
\eea
Hence we get
\be
\delta_{\Lambda}a_{\b}=\del_{\b}\lam
+g(\kappa^{\ds}\del_{\ds}a_{\b}+a_{\ds}\del_{\b}\kappa^{\ds}).
\label{transf-aa}
\ee

In our formulation of the self dual gauge field $b$, 
the components $b_{\mu\nu}$ do not explicitly show up in the action. 
Rather they appear when we solve the equations of motion for 
the rest of the components $b_{\dm\dn}$ and $b_{\mu\dm}$.
In \cite{Pasti:2009xc,Furuuchi:2010sp}, 
the components $b_{\mu\nu}$ are used to explicitly exhibit 
the self duality of the gauge field,
and their gauge transformation laws are given by
\be
\delta_{\Lambda}b_{\m\n}=\del_{\m}\Lambda_{\n}-\del_{\n}\Lambda_{\m}
+g[\kappa^{\dr}(\del_{\dr}b_{\m\n})
+(\del_{\n}\kappa^{\dr})b_{\m\dr}-(\del_{\m}\kappa^{\dr})b_{\n\dr}].
\ee
Identifying $b_{\b 2}$ with $a_{\b}$ and setting $\del_2 = 0$ for DDR,
we get exactly the same gauge transformation rule as (\ref{transf-aa})
with $\Lambda_2 = \lam$.

We find that the gauge transformation of $a_{\dm}$ (\ref{transf-adm}) 
and that of $a_{\a}$ (\ref{transf-aa}) are of the same form
\be
\delta_{\Lambda} a_{A}=
\del_{A}\lambda+g(\kappa^{\dn}\del_{\dn}a_{A}+a_{\dn}\del_{A}\kappa^{\dn}).
\label{transf-a}
\ee

For the convenience of the reader, 
let us also give here 
the gauge transformation of $V_{\dn}{}^{\dm}$, 
$M_{\dm\dn}{}^{\a\b}$ and $\hat{B}_{\a}^{~\dm}$:
\bea
\delta_{\Lambda} V_{\dn}{}^{\dm} &=&
g\kappa^{\dlam}\del_{\dlam}V_{\dn}{}^{\dm} 
+ g(\del_{\dn}\kappa^{\dlam}) V_{\dlam}{}^{\dm},
\\
\delta_{\Lambda} M_{\dm\dn}{}^{\a\b} &=&
g[\kappa^{\ds}\del_{\ds}M_{\dm\dn}{}^{\a\b}
+(\del_{\dm}\kappa^{\ds})M_{\ds\dn}{}^{\a\b}
+(\del_{\dn}\kappa^{\ds})M_{\dm\ds}{}^{\a\b}], 
\\
\delta_{\Lambda} \hat{B}_{\a}^{~\dm} &=& 
\del_{\a}\kappa^{\dm}+
g(\kappa^{\dn}\del_{\dn}\hat{B}_{\a}^{~\dm}-
\hat{B}_{\a}^{~\dn}\del_{\dn}\kappa^{\dm}).
\label{transf-B} 
\eea

\subsection{Covariant Field Strengths}

In the original NP M5-brane theory, 
we have the covariant field strengths
\footnote{
A field $\Phi$ is covariant if its gauge transformation is 
$\delta_{\Lambda} \Phi = g\kappa^{\dm}\del_{\dm}\Phi$.
}
\bea
{\cal H}_{\dot{1}\dot{2}\dot{3}}&=& 
\del_{\dm}b^{\dm}+\frac{1}{2}g
(\del_{\dn}b^{\dn}\del_{\dr}b^{\dr}-\del_{\dn}b^{\dr}\del_{\dr}b^{\dn})
+g^2 \{b^{\dot{1}},b^{\dot{2}},b^{\dot{3}}\}, \\
{\cal F}_{\dm\dn}&\equiv& {\cal H}_{2\dm\dn} =
F_{\dm\dn}+g
[\del_{\ds}b^{\ds}F_{\dm\dn}-\del_{\dm}b^{\ds}F_{\ds \dn}
-\del_{\dn}b^{\ds}F_{\dm\ds}],
\eea
which survive the DDR.
Here we have also rewritten ${\cal H}_{2\dm\dn}$, 
which was given above in (\ref{H2dmdn}), 
in a different but equivalent form.

The covariant version of $F_{\a\dm}$ can be defined as
\be
{\cal F}_{\a\dm} \equiv 
\frac{1}{2}\eps_{\b\a}\eps_{\dm\dn\dlam}{\cal H}^{\b\dn\dlam}. 
\ee
This is motivated by the intuition that ${\cal F}_{\a\dm}$ 
corresponds to ${\cal H}_{\a\dm 2}$ in the M5-brane theory, 
and we used the self duality condition of ${\cal H}$
to write down the expression above.
Replacing $B_{\a}{}^{\dm}$ by the solution $\hat{B}_{\a}{}^{\dm}$, 
we can rewrite ${\cal H}^{\b\dn\dlam}$ (\ref{Hadmdn}) as
a function of $F_{AB}$, $\del_{\dm} b^{\dn}$ and $\hat{B}_{\a}{}^{\dm}$.
(That is, we avoided using $\del_{\a} b^{\dm}$ directly.
The dependence on $\del_{\a} b^{\dm}$ only appears through $\hat{B}_{\a}{}^{\dm}$.)
As a result, we have
\be
{\cal F}_{\a\dm} =
V^{-1}{}_{\dm}{}^{\dn}(F_{\a\dn} + gF_{\dn\ds}\hat{B}_{\a}{}^{\ds}).
\ee
This is also in agreement with the definition of ${\cal H}_{\mu\nu\dm}$
defined in \cite{Pasti:2009xc,Furuuchi:2010sp}.

By inspection, 
we can guess the covariant form of $F_{\a\b}$.
Together with the rest of the covariant field strengths of the $U(1)$ gauge field, 
we have
\bea
{\cal F}_{\dm\dn}&=&
F_{\dm\dn}+g[\del_{\ds}b^{\ds}F_{\dm\dn}-
\del_{\dm}b^{\ds}F_{\ds \dn}-\del_{\dn}b^{\ds}F_{\dm \ds}]
\nn \\
&=&V^{~\dr}_{\dr}F_{\dm\dn}+V^{~\dr}_{\dm}F_{\dn\dr}+
V^{~\dr}_{\dn}F_{\dr\dm}, 
\label{Fdmdn} \\
{\cal F}_{\a\dm}&=&
{V^{-1}}_{\dm}^{~\dn}(F_{\a\dn}+gF_{\dn\dd}\hat{B}_{\a}^{~\dd}), 
\label{Fadm} \\
{\cal F}_{\a\b}&=&
F_{\a\b}+g[-F_{\a\dm}\hat{B}_{\b}^{~\dm}-
F_{\dm\b}\hat{B}_{\a}^{~\dm}+
gF_{\dm\dn}\hat{B}_{\a}^{~\dm}\hat{B}_{\b}^{~\dn}],
\label{Fab}
\eea
where 
\be
F_{AB} \equiv \del_A a_B - \del_B a_A.
\ee
Unlike ${\cal F}_{\dm\dn}$ and ${\cal F}_{\a\dm}$,
the components ${\cal F}_{\a\b}$ can not be directly 
matched with the field ${\cal H}_{\a\b 2}$ in the M5-brane theory, 
because the latter involves other fields that 
does not exist in the D4-brane theory.

\subsection{D4-brane Action in Terms of Covariant Variables}

Remarkably, in terms of the covariant field strengths,
the action is simply
\be
S'_{gauge}[b^{\dm},a_{A}]=
\int d^{2}x d^{3}y \left\{
-\frac{1}{2}{\cal H}_{\dot{1}\dot{2}\dot{3}} {\cal H}^{\dot{1}\dot{2}\dot{3}}
-\frac{1}{4}{\cal F}_{\dn\dr}{\cal F}^{\dn\dr}
+\frac{1}{2}{\cal F}_{\b\dm}{\cal F}^{\b\dm}
+\frac{1}{2g}\eps^{\a\b}{\cal F}_{\a\b} \right\}.
\label{Sgauge1}
\ee
The last term in the Lagrangian resembles the Wess-Zumino term 
for the $C$-field.

It appears that we are missing the kinetic term 
${\cal F}_{\a\b}{\cal F}^{\a\b}$ in the Lagrangian,
and the coefficient of the term ${\cal F}_{\a\dm}{\cal F}^{\a\dm}$ is wrong.
However, in the next section we will see that 
the missing kinetic term arises when we integrate out $b^{\dm}$.

\section{D4-brane Action Expanded}
\label{ExpandD4}

\subsection{Zeroth Order}

In this subsection we show that at the lowest order of $g$,
the D4-brane action (\ref{Sgauge1}) agrees with
the Maxwell action for a $U(1)$ gauge field in
the ordinary D4-brane action.
First we expand everything to the 1st order
\bea
{\cal H}_{\dot{1}\dot{2}\dot{3}}&=&\del_{\dm}b^{\dm}+
g\frac{1}{2}(\del_{\dn}b^{\dn}\del_{\dr}b^{\dr}-\del_{\dn}b^{\dr}\del_{\dr}b^{\dn})
+ {\cal O}(g^2),
\\
{V^{-1}}_{\dm}^{~\dn}&=&\delta_{\dm}^{~\dn}-g\del_{\dm}b^{\dn}+O(g^2),
\\
(M^{-1})^{\dm\dn}_{~~~\a\b}&=&
\d^{\dm\dn}\d_{\a\b}-
g[(\del^{\dm}b^{\dn}+\del^{\dn}b^{\dm})\d_{\a\b}-\eps_{\a\b}F^{\dm\dn}]
+O(g^2),
\\
\hat{B}_{\a}^{~\dm}&=&\del_{\a}b^{\dm}+\eps_{\a\b}F^{\b\dm} 
+g[-\del^{\ds}b^{\dm}\del_{\a}b_{\ds}-\del^{\dm}b_{\ds}\eps_{\a\b}F^{\b\ds}
-\del_{\ds}b^{\dm}\eps_{\a\b}F^{\b\ds} \nn \\
&&+\eps_{\a\b}\del^{\b}b_{\ds}F^{\dm\ds}+F_{\a\ds}F^{\dm\ds}]+O(g^2).
\\
{\cal F}_{\b\dm}&=&F_{\b\dm}+
g(\del_{\dm}b^{\ds}F_{\ds\b}+\del_{\b}b^{\ds}F_{\dm\ds}+
\eps_{\b\g}F_{\dm\ds}F^{\g\ds})+O(g^2)
\\
{\cal F}_{\a\b}&=&F_{\a\b}+
g[-F_{\a\dm}(\del_{\b}b^{\dm}+\eps_{\b\g}F^{\g\dm})-
F_{\dm\b}(\del_{\a}b^{\dm}+\eps_{\a\g}F^{\g\dm})]+O(g^2).
\eea

To the lowest order of $g$, the last term in the Lagrangian (\ref{Sgauge1}) is
\bea
\frac{1}{2g}\eps^{\a\b}{\cal F}_{\a\b}
&=&\frac{1}{2g}\eps^{\a\b}F_{\a\b} +
\frac{1}{2}\eps^{\a\b}[-F_{\a\dm}(\del_{\b}b^{\dm}+\eps_{\b\g}F^{\g\dm})-
F_{\dm\b}(\del_{\a}b^{\dm}+\eps_{\a\g}F^{\g\dm})] + {\cal O}(g) \nn \\
&\simeq& - \eps^{\a\b}F_{\a\dm}\del_{\b}b^{\dm} - F_{\a\dm}F^{\a\dm}
+{\cal O}(g) \nn \\
&\simeq& \eps^{\a\b}\del_{\b}a_{\a}\del_{\dm}b^{\dm} - F_{\a\dm}F^{\a\dm}
+{\cal O}(g),
\eea
up to total derivatives.
To the 0-th order of $g$,
the action (\ref{Sgauge1}) can now be expressed as
\bea
S'{}^{(0)}_{gauge}[b^{\dm},a_{A}]&=&
\int d^{2}x d^{3}y \left\{
-\frac{1}{2}H_{\dot{1}\dot{2}\dot{3}}^2
- \frac{1}{2} \eps^{\a\b}F_{\a\b} H_{\dot{1}\dot{2}\dot{3}}
-\frac{1}{4}F_{\dm\dn}F^{\dm\dn}
-\frac{1}{2}F_{\a\dm}F^{\a\dm}
\right\} \nn \\
&=&
\int d^{2}x d^{3}y \left\{
-\frac{1}{2}(H_{\dot{1}\dot{2}\dot{3}}+F_{01})^2
-\frac{1}{4}F_{AB}F^{AB}
\right\},
\label{Sgauge10}
\eea
where $H_{\dot{1}\dot{2}\dot{3}} = \del_{\dm}b^{\dm}$ and $A,B=(\dm,\a)$.
Note that $H_{\dot{1}\dot{2}\dot{3}}$ is 
the only gauge invariant degree of freedom 
in the gauge potential $b^{\dm}$ 
because there are two independent gauge transformation parameters.
\footnote{
Since the 3 gauge transformation parameters $\kappa^{\dm}$
are subject to the condition $\del_{\dm}\kappa^{\dm}=0$,
there are only 2 functionally independent degrees of freedom in $\kappa^{\dm}$.
}
Furthermore there is no kinetic term for $b^{\dm}$ 
and so we can integrate it out and then
(\ref{Sgauge10}) becomes exactly the Maxwell action.
Integrating out $b^{\dm}$ is a duality transformation 
which imposes the identification
\be
H_{\dot{1}\dot{2}\dot{3}} = - F_{01}.
\ee
The physical degrees of freedom in $b^{\dm}$ is 
transformed into that of $a_{\a}$.
Although $b^{\dm}$ appears as new gauge potentials in 
the D4-brane theory, 
they share the same physical degrees of freedom with $a_{\a}$.

\subsection{First Order}
\label{FirstOrder}

The first order correction to the action (\ref{Sgauge10}) is
\bea
S'{}^{(1)}_{gauge}[b^{\dm},a_{A}] &=& 
g \int d^{2}x d^{3}y \left\{
(- \frac{1}{2}H_{\dot{1}\dot{2}\dot{3}}^2+\frac{1}{2}\del_{\dm}b^{\dn}\del_{\dn}b^{\dm})
(H_{\dot{1}\dot{2}\dot{3}} + F_{01})\right.\nn\\
&&+ H_{\dot{1}\dot{2}\dot{3}}
\left(
- \frac{1}{2}F_{\dm\dn}F^{\dm\dn} 
+ \eps^{\a\b}F_{\a\dm}\del_{\b}b^{\dm}\right)- \frac{1}{2}\eps_{\a\b}F_{\dm\dn}F^{\a\dm}F^{\b\dn}
\nn \\
&& \left.
+ F^{\dm\dn}F_{\dlam\dn}\del_{\dm}b^{\dlam}
+ F_{\a\dm}F^{\a}{}_{\dn}\del^{\dm}b^{\dn}
- F_{\a\dm}\del^{\a}b_{\dn}F^{\dm\dn}
\right\}.
\eea

In order to integrate out $H_{\dot{1}\dot{2}\dot{3}}$, 
note that we can impose the gauge fixing condition 
\be
\eps^{\dm\dn\dlam}\del_{\dm}b_{\dn} = 0,
\ee
so that 
\be
b_{\dm} = \del_{\dm}c
\ee
for some function $c$.
Solving $c$ from
\be
H_{\dot{1}\dot{2}\dot{3}} = \del_{\dm}b^{\dm},
\label{Hdb}
\ee
we find
\be
b^{\dm} = \del^{\dm}\dot{\del}^{-2}H_{\dot{1}\dot{2}\dot{3}},
\label{b-H}
\ee
where $\dot{\del}^{-2}$ is the inverse operator 
of the Laplacian $\dot{\del}^2 \equiv \del_{\dm}\del^{\dm}$.
Denoting the Green's function of the Laplacian by $G$ so that
\be
\dot{\del}^2 G(y-y') = \delta^{(3)}(y-y'),
\ee
where $y$ and $y'$ represent the coordinates 
in the directions $y^{\dot{1}}, y^{\dot{2}}, y^{\dot{3}}$.
We have
\be
\dot{\del}^{-2}\phi(y) = \int d^3 y' \; G(y-y') \phi(y).
\ee

Plugging (\ref{b-H}) into the action, 
we get an action as a functional of $H_{\dot{1}\dot{2}\dot{3}}$ and $a_A$.
To the first order in $g$, 
we can integrate out $H_{\dot{1}\dot{2}\dot{3}}$
and the action becomes
\bea
S''_{gauge}[a_{A}] &=&
\int d^{2}x d^{3}y \left\{
-\frac{1}{4}F_{AB}F^{AB}
+ g\left[
- F_{01} {\cal C}
- \frac{1}{2}\eps_{\a\b}F_{\dm\dn}F^{\a\dm}F^{\b\dn}
\right. \right. \nn \\
&& \left. \left.
- F^{\dm\dn}F_{\dlam\dn}\del_{\dm}\del^{\dlam}\dot{\del}^{-2}F_{01}
- F_{\a\dm}F^{\a}{}_{\dn}\del^{\dm}\del^{\dn}\dot{\del}^{-2}F_{01}
+ F_{\a\dm}F^{\dm\dn}\del^{\a}\del_{\dn}\dot{\del}^{-2}F_{01}
\right]
\right\},
\eea
where
\bea
{\cal C} &=& 
- \frac{1}{2}F_{\dm\dn}F^{\dm\dn}- 
\eps^{\a\b}F_{\a\dm}\del_{\b}\del^{\dm}\dot{\del}^{-2}F_{01}.
\eea
Apparently the action becomes nonlocal at order ${\cal O}(g)$.

In principle, using (\ref{b-H}) to rewrite the action 
as a functional of $a_A$ and $H_{\dot{1}\dot{2}\dot{3}}$,
we can integrate out $H_{\dot{1}\dot{2}\dot{3}}$ to an arbitrary order in $g$.
The resulting action would be a functional of $F_{AB}$
with higher derivatives and $\dot{\del}^{-2}$.

\section{Matter Fields}
\label{Matter}

In the above we have ignored the matter fields in the NP M5-brane theory.
It is straightforward to repeat the derivations above
with the matter fields included. 
Analogous to (\ref{S4}), we get
\bea
S^{(4)}[b^{\dm},a_A,\breve{B}_{\a}^{~\dm},X^{i},\Psi]&=&
\int d^{2}x d^{3}y\left\{
-\frac{1}{2}{\cal D}_{\dm}X^{i}{\cal D}^{\dm}X^{i} -\frac{1}{2}\del_{\a}X^{i}\del^{\a}X^{i}+g\breve{B}_{\a}^{~\dm}\del_{\dm}X^{i}\del^{\a}X^{i} \right.
\nn\\
&&- \frac{g^{2}}{2}\breve{B}_{\a}^{~\dm}\breve{B}^{\a}_{~\dn}
\del_{\dm}X^{i}\del^{\dn}X^{i} 
-\frac{g^{2}}{8}\eps^{\dm\dr\dt}\eps_{\dn\ds\dd}F_{\dr\dt}F^{\ds\dd}
\del_{\dm}X^{i}\del^{\dn}X^{i}
\nn\\
&&-\frac{g^{4}}{4}\{X^{\dm},X^{i},X^{j}\}^{2}
-\frac{g^{4}}{12}\{X^{i},X^{j},X^{k}\}^{2}
\nn \\
&&+\frac{i}{2}\bar{\Psi}\Gamma^{\a}\del_{\a}\Psi
+\frac{i}{2}\bar{\Psi}\Gamma^{\dr}{\cal D}_{\dr}\Psi 
+g\frac{i}{4}\bar{\Psi}\Gamma^{2}\eps^{\dm\dn\dr}F_{\dn\dr}\del_{\dm}\Psi 
-g\frac{i}{2}\bar{\Psi}\Gamma^{\a}\breve{B}_{\a}^{~\dm}\del_{\dm}\Psi
\nn\\
&&+g^2\frac{i}{2}\bar{\Psi}\Gamma_{\dm i}\{X^{\dm},X^{i},\Psi\}
+g^2\frac{i}{4}\bar{\Psi}\Gamma_{ij}
\Gamma_{\dot{1}\dot{2}\dot{3}}\{X^{i},X^{j},\Psi\}
\nn\\
&&-\frac{1}{2g^{2}} -\frac{1}{2}({\cal H}_{\dot{1}\dot{2}\dot{3}})^{2}
-\frac{1}{4}{\cal F}_{\dn\dr}{\cal F}^{\dn\dr}
-\frac{1}{4}(\eps_{\dm\dn\dr}
(\del_{\a}b^{\dm}-V_{\ds}^{~\dm}\breve{B}_{\a}^{~\ds}))^2
\nn \\
&&\left. 
+\eps^{\a\b}F_{\b\dm}\breve{B}_{\a}^{~\dm}
+\frac{g}{2}\eps^{\a\b}F_{\dm\dn}\breve{B}_{\a}^{~\dm}\breve{B}_{\b}^{~\dn}
\right\}.
\eea


With the matter fields included,
the action is still no more than quadratic in $\breve{B}_{\a}^{~\dm}$
and so we can still integrate it out.
This is equivalent to solving the equation of motion for $\breve{B}_{\a}^{~\dm}$
and plugging it back into the action.
The new equation of motion for $\breve{B}_{\a}^{~\dm}$ is
\be
V_{\dm}^{~\dn}(\del^{\a}b_{\dn}-V^{\dr}_{~\dn}\breve{B}^{\a}_{~\dr})
+\eps^{\a\b}F_{\b\dm}+g\eps^{\a\b}F_{\dm\dn}\breve{B}_{\b}^{~\dn}
+g\del_{\dm}X^{i}\del^{\a}X^{i}-g\frac{i}{2}\bar{\Psi}\Gamma^{\a}\del_{\dm}\Psi
-g^{2}\breve{B}^{\a}_{~\dn}\del_{\dm}X^{i}\del^{\dn}X^{i}=0.
\label{r2}
\ee
Its solution is
\bea
\hat{B}_{\a}^{~\dm}&=& (\textbf{M}^{-1})^{\dm\dn}_{~~~\a\b}
(V_{\dn}^{~\ds}\del^{\b}b_{\ds}+\eps^{\b\g}F_{\g\dn}
+g\del_{\dn}X^{i}\del^{\b}X^{i}-g\frac{i}{2}\bar{\Psi}\Gamma^{\b}\del_{\dn}\Psi) 
\nn \\
&\equiv&(\textbf{M}^{-1})^{\dm\dn}_{~~~\a\b}W_{\dn}^{\b},
\eea
where
\be
\textbf{M}_{\dm\dn}^{~~\a\b} \equiv
(V_{\dm\dr}V_{\dn}^{~\dr}+g^{2}\del_{\dm}X^{i}\del_{\dn}X^{i})\d^{\a\b}
-g\eps^{\a\b}F_{\dm\dn},
\ee
and $(\textbf{M}^{-1})^{\dm\dn}_{~~~\a\b}$ is defined by
\be
(\textbf{M}^{-1})^{\dlam\dm}_{~~~\g\a}\textbf{M}_{\dm\dn}^{~~\a\b}
=\d^{\dlam}_{~~\dn}\d^{~\b}_{\g}.
\ee

Finally, we get the action 
\bea
S'_{gauge}[b^{\dm},a_{A},X^{i},\Psi]&=&
\int d^{2}x d^{3}y \left\{
-\frac{1}{2}{\cal D}_{\dm}X^{i}{\cal D}^{\dm}X^{i} 
-\frac{1}{2}\del_{\a}X^{i}\del^{\a}X^{i}
\right.\nn\\
&&-\frac{g^{2}}{8}\eps^{\dm\dr\dt}\eps_{\dn\ds\dd}F_{\dr\dt}F^{\ds\dd}\del_{\dm}
X^{i}\del^{\dn}X^{i}
\nn\\
&&-\frac{g^{4}}{4}\{X^{\dm},X^{i},X^{j}\}^{2}
-\frac{g^{4}}{12}\{X^{i},X^{j},X^{k}\}^{2}
\nn\\
&&+\frac{i}{2}\bar{\Psi}\Gamma^{\a}\del_{\a}\Psi
+\frac{i}{2}\bar{\Psi}\Gamma^{\dr}{\cal D}_{\dr}\Psi 
+g\frac{i}{4}\bar{\Psi}\Gamma^{2}\eps^{\dm\dn\dr}F_{\dn\dr}\del_{\dm}\Psi 
\nn\\
&&+g^2\frac{i}{2}\bar{\Psi}\Gamma_{\dm i}\{X^{\dm},X^{i},\Psi\}
+g^2\frac{i}{4}\bar{\Psi}\Gamma_{i j}\Gamma_{\dot{1}\dot{2}\dot{3}}
\{X^{i},X^{j},\Psi\} 
\nn\\
&&\left. -\frac{1}{2g^{2}} -\frac{1}{2}({\cal H}_{\dot{1}\dot{2}\dot{3}})^{2}
-\frac{1}{4}{\cal F}_{\dn\dr}{\cal F}^{\dn\dr}
+\frac{1}{2}W^{\a}_{\dm}(\textbf{M}^{-1})^{\dm\dn}_{~~\a\b}W^{\b}_{\dn}
\right\}.
\label{FullAction}
\eea
Here the fields ${\cal F}$ are defined by the same expressions 
as before but with the new definition of $\hat{B}$.




At the 0-th order of $g$, 
the action is just
\be
S''{}^{(0)}_{gauge}[a_{A},X^{i},\Psi] \simeq \int d^{2}x d^{3}y 
\left\{-\frac{1}{4}F_{AB}F^{AB}-\frac{1}{2}\del_{A}X^{i}\del^A X^{i}
+\frac{i}{2}\bar{\Psi}\Gamma^{A}\del_A\Psi\right\}
\ee
after we integrate out the VPD gauge fields.
This defines a Maxwell's theory with neutral bosons $X$ and fermions $\Psi$.


For completeness let us also give the expression of the action
to the 1st order:
\bea
S'_{gauge}[b^{\dm},a_A,X^{i},\Psi]&\simeq&\int d^{2}x d^{3}y 
\left\{-\frac{1}{2}\del_{\a}X^{i}\del^{\a}X^{i}
-\frac{1}{2}\del_{\dm}X^{i}\del^{\dm}X^{i} 
+\frac{i}{2}\bar{\Psi}\Gamma^{\a}\del_{\a}\Psi
+\frac{i}{2}\bar{\Psi}\Gamma^{\dm}\del_{\dm}\Psi
\right. \nn \\
&&
+g\eps^{\a\b}F_{\b\dm}\del^{\dm}X^{i}\del_{\a}X^{i}
+g\del^{\dm}X^{i}\del_{\a}X^{i}\del^{\a}b_{\dm}
-g\frac{i}{2}\eps^{\a\b}F_{\b\dm}\bar{\Psi}\Gamma_{\a}\del^{\dm}\Psi
\nn \\
&&
-g\frac{i}{2}\bar{\Psi}\Gamma_{\a}\del^{\dm}\Psi\del^{\a}b_{\dm}
-g\del_{\dm}X^{i}\del^{\dm}X^{i}\del_{\dr}b^{\dr}
+g\del^{\dm}X^{i}\del_{\dr}X^{i}\del_{\dm}b^{\dr}
\nn\\
&&
+g\frac{i}{2}\bar{\Psi}\Gamma^{\dr}\del_{\dr}\Psi\del_{\dn}b^{\dn}
-g\frac{i}{2}\bar{\Psi}\Gamma^{\dr}\del_{\dn}\Psi\del_{\dr}b^{\dn}
+g\frac{i}{4}\bar{\Psi}\Gamma^{2}\eps^{\dm\dn\dr}F_{\dn\dr}\del_{\dm}\Psi
\nn\\
&&
-\frac{1}{2}{\cal H}_{\dot{1}\dot{2}\dot{3}}{\cal H}^{\dot{1}\dot{2}\dot{3}}
-\frac{1}{4}{\cal F}_{\dn\dr}{\cal F}^{\dn\dr} 
-\frac{1}{2}F_{\b\dm}F^{\b\dm}-\frac{1}{2}\eps^{\a\b}F_{\a\b}\del_{\dm}b^{\dm}
\nn\\
&&
-g\eps^{\a\b}F_{\b\dm}\del_{\a}b_{\dn}\del^{\dn}b^{\dm}
+\frac{1}{2}g\eps^{\a\b}F_{\dm\dn}\del_{\a}b^{\dm}\del_{\b}b^{\dn}
+gF_{\dm\dn}F^{\a\dn}\del_{\a}b^{\dm}
\nn\\
&&
\left.
+gF^{\a\dn}F_{\a\dm}\del_{\dn}b^{\dm}
+\frac{1}{2}g\eps^{\a\b}F_{\b\dm}F^{\dm\dn}F_{\a\dn}+O(g^2)
\right\}.
\eea 


The full action (\ref{FullAction}) inherits the full supersymmetry from 
the NP M5-brane theory because DDR preserves global SUSY, 
and duality transformation is an equivalence relation.
Nevertheless it is not totally trivial to derive the explicit SUSY 
transformation rules for all the variables, 
in particular those arise as Lagrange multipliers. 
We leave this for future study.

\section{Generalization to Multiple D$p$-branes}
\label{Generalization}

To generalize the story about a single D4-brane to a system of multiple D$p$-branes, 
we notice first that 
the VPD for a volume $(p-1)$-form is generated by a $(p-1)$-bracket
\be
\{f_1, f_2, \cdots, f_{p-1}\} \equiv
\eps^{\dm_1 \dm_2 \cdots \dm_{p-1}} 
\del_{\dm_1}f_1 \del_{\dm_2}f_2 \cdots \del_{\dm_{p-1}}f_{p-1}.
\ee
We define a $(p-2)$-form gauge potential $b_{\dm_1\cdots\dm_{p-2}}$ 
and its dual
\be
b^{\dm_1} = \frac{1}{(p-2)!}\eps^{\dm_1 \dm_2 \cdots \dm_{p-1}} 
b_{\dm_2\cdots\dm_{p-1}}.
\ee
Let 
\be
X^{\dm} = \frac{y^{\dm}}{g} + b^{\dm}
\ee
and the field strength ${\cal H}$ can be defined as
\be
{\cal H}_{\dm_1\dm_2\cdots\dm_{p-1}} \equiv 
g^{p-2}\{X^{\dm_1}, X^{\dm_2}, \cdots, X^{\dm_{p-1}}\}-\frac{1}{g}
= \del_{\dm} b^{\dm} + {\cal O}(g).
\ee

In terms of $b^{\dm}$ the gauge transformation is exactly of 
the same form as (\ref{transf-bdm}), 
and the parameter $\kappa^{\dm}$ is still divergenceless.
The only change is that the range of the indices $\dm, \dn$ 
becomes $2, 3, \cdots, p$.
\footnote{
The indices $2, 3, \cdots$ would be denoted as $\dot{1}, \dot{2}, \cdots$ 
in previous sections.
}

While we do not intend to promote the VPD gauge potential $b^{\dm}$
to a matrix mostly because 
we do not know how to modify its gauge transformation law, 
we shall replace the $U(1)$ potential
by a $U(N)$ potential $a_A$, 
which is now an $N\times N$ anti-Hermitian matrix of 1-forms.
The $U(N)$ gauge transformation of $a_A$ should be defined by
\be
\d a_A = [D_A, \lam] + g(\kappa^{\dm}\del_{\dm}a_A 
+ a_{\dm} \del_A \kappa^{\dm}),
\label{transf-a-N}
\ee
where $D_A \equiv \del_A + a_A$.
It modifies (\ref{transf-a}) only by replacing $\del_A\lam$ 
by $[D_A, \lam]$.
The gauge transformation parameter $\lam$ 
is an $N\times N$ anti-Hermitian matrix
but $\kappa^{\dm}$ is $1\times 1$.
The range of the index $A$ is now
$A = 0, 1, 2, \cdots, p$.
Decomposing the potential $a_A$ into the $U(1)$ part and the $SU(N)$ part
\be
a_A = a_A^{U(1)} + a_A^{SU(N)},
\ee
the gauge transformation of $a_A^{U(1)}$ 
is exactly the same as before (\ref{transf-a}).

We can define $V_{\dm}{}^{\dn}$ and $\hat{B}_{\a}{}^{\dm}$ 
using the same expressions (\ref{def-V})--(\ref{def-B}) as before
\bea
V_{\dn}^{~\dm}&\equiv&\d_{\dn}^{~\dm}+g\del_{\dn}b^{\dm}, \\
M_{\dm\dn}{}^{\a\b} &\equiv&
V_{\dm\dr}V_{\dn}{}^{\dr}\d^{\a\b}-g\eps^{\a\b}F^{U(1)}_{\dm\dn},
\\
\hat{B}_{\a}^{~\dm} &\equiv&
(M^{-1})^{\dm\dn}{}_{\a\b}
(V_{\dn}^{~\ds}\del^{\b}b_{\ds}+\eps^{\b\g}F^{U(1)}_{\g\dn}),
\eea
but with the field strength $F^{U(1)}_{\dm\dn}$ being 
the $U(1)$ part of the $U(N)$ field strength,
so that their gauge transformations remain the same.
The range of the indices $\a, \b$ is still $0, 1$.

The naive definition of field strength $F_{AB} \equiv [D_A, D_B]$ is not covariant. 
They transform like
\be
\delta F_{AB} =
[F_{AB}, \lam] + g\kappa^{\dm}\del_{\dm}F_{AB}
+ g[ (\del_A \kappa^{\dm}) F_{\dm B} - (\del_B \kappa^{\dm}) F_{\dm A} ].
\ee
It turns out that exactly the same expressions as (\ref{Fdmdn})--(\ref{Fab})
give the covariant field strengths.
For the convenience of the reader we reproduce them here
\bea
{\cal F}_{\dm\dn}&=&
F_{\dm\dn}+g[\del_{\ds}b^{\ds}F_{\dm\dn}-
\del_{\dm}b^{\ds}F_{\ds \dn}-\del_{\dn}b^{\ds}F_{\dm \ds}]
\nn \\
&=&V^{~\dr}_{\dr}F_{\dm\dn}+V^{~\dr}_{\dm}F_{\dn\dr}+
V^{~\dr}_{\dn}F_{\dr\dm}, 
\label{Fdmdn1} \\
{\cal F}_{\a\dm}&=&
{V^{-1}}_{\dm}^{~\dn}(F_{\a\dn}+gF_{\dn\dd}\hat{B}_{\a}^{~\dd}), 
\label{Fadm1} \\
{\cal F}_{\a\b}&=&
F_{\a\b}+g[-F_{\a\dm}\hat{B}_{\b}^{~\dm}-
F_{\dm\b}\hat{B}_{\a}^{~\dm}+
gF_{\dm\dn}\hat{B}_{\a}^{~\dm}\hat{B}_{\b}^{~\dn}].
\label{Fab1}
\eea
They transform like
\be
\d {\cal F}_{AB} = [{\cal F}_{AB}, \lam - g\kappa^{\dm}\del_{\dm}].
\ee
From this expression it is easy to check that
the gauge symmetry algebra is given by
\be
[\d_1, \d_2] = \d_3,
\ee
where $\d_i$ is the gauge transformation 
with parameters $\lam_i, \kappa^{\dm}_i$ and
\bea
\lam_3 &=& [\lam_1, \lam_2] + 
g(\kappa_2^{\dm}\del_{\dm}\lam_1 - \kappa_1^{\dm}\del_{\dm}\lam_2),
\\
\kappa_3^{\dm} &=&
g(\kappa_2^{\dn}\del_{\dn}\kappa_1^{\dm}
-\kappa_1^{\dn}\del_{\dn}\kappa_2^{\dm}).
\eea

In view of the D4-brane action (\ref{Sgauge1}),
it is now natural to define the action for the gauge fields 
on multiple D$p$-branes in R-R $(p-1)$-form field background as
\bea
S^{Dp}_{gauge}[b^{\dm},a_{A}] &=&
\int d^{2}x d^{p-1}y \left\{
-\frac{1}{2}\frac{1}{(p-1)!}
{\cal H}_{\dm_1\cdots\dm_{p-1}} {\cal H}^{\dm_1\cdots\dm_{p-1}}
+\frac{1}{2g}\eps^{\a\b}{\cal F}_{\a\b}^{U(1)} \right.
\nn \\
&& 
\left.
-\frac{1}{4}{\cal F}^{U(1)}_{\dn\dr}{\cal F}_{U(1)}^{\dn\dr}
+\frac{1}{2}{\cal F}^{U(1)}_{\b\dm}{\cal F}_{U(1)}^{\b\dm}
-\frac{1}{4}\mbox{tr}\left({\cal F}^{SU(N)}_{AB}{\cal F}_{SU(N)}^{AB}\right)
\right\}.
\label{SgaugeDp}
\eea
If we focus our attention on the $U(1)$ part of the 1-form gauge potential $a_A$
and the VPD gauge potential $b^{\dm}$, 
everything is exactly the same as before. 
The VPD field strength ${\cal H}$ is dual to only the $U(1)$ part of ${\cal F}_{01}$.
But since the $SU(N)$ part of the field strength ${\cal F}_{AB}$ 
involves the VPD potential $b^{\dm}$, 
the $U(1)$ part of $a_A$ couples to the $SU(N)$ part indirectly through $b^{\dm}$.
This is different from the usual Yang-Mills theory of $U(N)$ gauge symmetry, 
for which the $U(1)$ part decouples, 
but similar to the noncommutative $U(N)$ YM theory.

To the 0-th order in $g$, 
the action is
\be
S'{}^{Dp(0)}_{gauge}[b^{\dm},a_{A}]=
\int d^{2}x d^{3}y \left\{
-\frac{1}{2}(H_{23\cdots p}+F^{U(1)}_{01})^2
-\frac{1}{4}F^{U(1)}_{AB}F_{U(1)}^{AB}
-\frac{1}{4}\mbox{tr}\left(F^{SU(N)}_{AB}F_{SU(N)}^{AB}\right)
\right\},
\label{Sgauge10Dp}
\ee
where $H_{23\cdots p} = \del_{\dm}b^{\dm}$.
Again, since $H_{23\cdots p}$ is the only component of the field strength 
for the gauge potential $b^{\dm}$, 
we can integrate it out and the action reduces to 
that of a Yang-Mills theory in $(p+1)$ dimensions.

In fact, the VPD symmetry allows us to impose 
the gauge fixing condition 
\be
\del^{\dm_1}b_{\dm_1\dm_2\cdots\dm_{p-2}} = 0 
\qquad \Leftrightarrow \qquad
\del^{\dm}b^{\dn}-\del^{\dn}b^{\dm} = 0.
\ee
This condition allows us to solve $b^{\dm}$ in terms of $H_{23\cdots p}$ as 
\be
b^{\dm} = \del^{\dm} \dot{\del}^{-2} H_{23\cdots p},
\label{bfromH}
\ee
where $\dot{\del}^{-2}$ is the inverse of the Laplace operator 
$\dot{\del}^2 \equiv \del_{\dm}\del^{\dm}$.
Like what we did in Sec. \ref{FirstOrder}, 
we can continue to integrate out $H_{23\cdots p}$ at higher orders of $g$
using the relation (\ref{bfromH}) for every term involving $b^{\dm}$ in the action, 
although we would get a nonlocal action in the end.
In principle we can write down a nonlocal action of $a_A$ 
without any trace of $b^{\dm}$ or $H_{23\cdots p}$
as an expansion of $g$ to an arbitrary order.

\section{Conclusion and Discussion}
\label{Conclusion}

In this paper, we showed that the $C$-field background 
induces a Nambu-Poisson structure on the D4-brane which generates
the gauge symmetry of volume-preserving diffeomorphisms.
The potential $b^{\dm}$ for this new gauge symmetry 
is the electric-magnetic dual of the $U(1)$ gauge potential $a_{\a}$
in the D4-brane theory.
In the limit $g\rightarrow 0$ the NP D4-brane theory reduces to 
the usual Maxwell theory for a D4-brane.
Our D4-brane action in $C$-field background is a good approximation 
of the system in the limit (\ref{D4limit}) and (\ref{gl}).

It is possible that the NP M5-brane theory is good at low energy 
for a wider range of limits than that given in (\ref{NPlimit}).
It may be possible that it is a good effective theory when 
$g_{22} \sim \eps^{2a}$ for some positive real number $a$ 
within a certain range,
while all other variables take the same limit as before. 
This would allow us to impose the condition (\ref{gl}) simultaneously with (\ref{D4limit}).
The compactification radius $R$ 
in the double dimensional reduction scales like
\be 
R \sim \eps^a, 
\ee
if $g_{22} \sim \eps^{2a}$.
The type \IIA parameters then scale like
\be
\ell_s \sim \eps^{(1-a)/2}, \qquad
g_s \sim \eps^{(3a-1)/2},
\ee
while $2\pi \a' B_{01}$ is still finite when $\eps \rightarrow 0$.
If
\be
1/3 < a < 1,
\ee
we would have $\ell_s, g_s, R/\ell_s, R/\ell_P \rightarrow 0$ 
at the same time.

In the previous section we have partially generalized the NP D4-brane theory 
to theories for D$p$-branes in constant $(p-1)$-form background, 
although we have not yet included matter fields.
In the gauge field sector of the D$p$-brane theory,
there is a $(p-2)$-form gauge potential $b$
for the volume-preserving diffeomorphism symmetry 
in the $(p-1)$ directions selected by the R-R $(p-1)$-form background.
The volume-preserving diffeomorphism is generated by 
a generalization of the Nambu-Poisson bracket with $(p-1)$ slots.
The field strength ${\cal H}$ of the potential $b^{\dm}$ is dual to 
the $U(1)$ field strength in the sense that, 
to the lowest order in $g$,
\be
F^{01} = \frac{1}{(p-1)!}\eps^{01\dm_1\cdots\dm_{p-1}}
H_{\dm_1\cdots\dm_{p-1}},
\ee
so there is no independent physical degrees of freedom 
in $b^{\dm}$ apart from those in the $U(1)$ gauge field $a$.
The all-order relation between the covariant field strength ${\cal H}$
and ${\cal F}$ is very complicated
because the volume preserving diffeomorphism is non-Abelian.
Both the definitions of ${\cal H}$ and ${\cal F}$ are quite non-trivial.
Apart from the advancement in our understanding about D-branes 
in string theory, 
the novelty of the structure of gauge symmetry 
is intriguing by itself.
This is an efficient gauge theory in which the gauge potentials for
two gauge symmetries share the same physical degrees of freedom.

If we take T-duality along the $x^1$-direction of the D4-brane 
considered in this paper, 
we get a D3-brane in an R-R 4-form potential background. 
The background 4-form can be decomposed as the wedge product of 
a 1-form in the direction $\tilde{x}^1$ T-dual to $x^1$ 
and the 3-form $C$ along the D3-brane worldvolume.
The 3-form $C$ defines a volume-form and the corresponding VPD 
is inherited from the D4-brane.
We still need the gauge fields $b^{\dm}$ for D3.
The connection between $F_{01}$ and $H_{\dot{1}\dot{2}\dot{3}}$ 
at the 0-th order for a D4-brane becomes 
the connection between $\tilde{p}_1$, 
the momentum in the direction of $\tilde{x}^1$, 
and $H_{\dot{1}\dot{2}\dot{3}}$.
Extending this conclusion to D$p$-branes, 
we claim that for a D$p$-brane in the RR $(p+1)$-form potential background 
\be
D^{(p+1)} = V^{(1)} \wedge C^{(p)}, 
\ee
where $V^{(1)}$ is transverse and $C^{(p)}$ is parallel to the D$p$-brane,
the VPD corresponding to the volume-form $C^{(p)}$ 
shares the same gauge field degrees of freedom with 
the component of the momentum $p$ in the direction of $V^{(1)}$.

For future works, we would like to include matter fields 
for D$p$-branes in large R-R $(p-1)$-form gauge potential background.
It will also be important to find the explicit expressions 
of SUSY transformation laws.

As a final goal of this line of research, 
one would like to generalize the results to the ultimate generality 
of multiple D$p$-branes and NS 5-branes in all combinations
NS-NS and R-R field background in various limits.
An immediate challenging problem is to 
define a deformation of VPD such that 
it is the electric magnetic dual of 
the noncommutative and/or non-Abelian gauge symmetry.
There are related No-Go theorems \cite{Bekaert:1999dp,Chen:2010ny}
suggesting that
there will be brand new gauge symmetries yet to be discovered.

\section*{Acknowledgment}

The authors thank Chien-Ho Chen, Wei-Ming Chen, Chong-Sun Chu, 
Kazuyuki Furuuchi, Petr Ho\v{r}ava, Kuo-Wei Huang, Yu-tin Huang,
Hirotaka Irie, Hiroshi Isono, Sheng-Lan Ko, Yutaka Matsuo, Yu Nakayama,
Hirosi Ooguri, Ryu Sasaki, John Schwarz, Tomohisa Takimi, Wen-Yu Wen 
and Chen-Pin Yeh for helpful discussions. 
The final stage of this work was completed during a visit of P.M.H. at CalTech.
He thanks the hospitality of the high energy theory group at CalTech.
This work is supported in part by
the National Science Council,
the NSC internship program,
the National Center for Theoretical Sciences, 
and the Center for Theoretical Sciences at National Taiwan University.


\vskip .8cm
\baselineskip 22pt
\begin{thebibliography}{99}
\itemsep 0pt

\bibitem{Leigh}
  R.~G.~Leigh,
  ``Dirac-Born-Infeld Action from Dirichlet Sigma Model,''
  Mod.\ Phys.\ Lett.\  A {\bf 4}, 2767 (1989).
  
\bibitem{Witten:1995im}
  E.~Witten,
  ``Bound states of strings and p-branes,''
  Nucl.\ Phys.\  B {\bf 460}, 335 (1996)
  [arXiv:hep-th/9510135].

\bibitem{CH}
  C.~S.~Chu and P.~M.~Ho,
  ``Noncommutative open string and D-brane,''
  Nucl.\ Phys.\  B {\bf 550}, 151 (1999)
  [arXiv:hep-th/9812219].
  C.~S.~Chu and P.~M.~Ho,
  ``Constrained quantization of open string in background B field and
  noncommutative D-brane,''
  Nucl.\ Phys.\  B {\bf 568}, 447 (2000)
  [arXiv:hep-th/9906192].

\bibitem{Schomerus}
  V.~Schomerus,
  ``D-branes and deformation quantization,''
  JHEP {\bf 9906}, 030 (1999)
  [arXiv:hep-th/9903205].

\bibitem{Seiberg:1999vs}
  N.~Seiberg and E.~Witten,
  ``String theory and noncommutative geometry,''
  JHEP {\bf 9909}, 032 (1999)
  [arXiv:hep-th/9908142].
  
\bibitem{oldM5-1}
  P.~S.~Howe and E.~Sezgin,
  ``D = 11, p = 5,''
  Phys.\ Lett.\  B {\bf 394}, 62 (1997)
  [arXiv:hep-th/9611008].
  P.~S.~Howe, E.~Sezgin and P.~C.~West,
  ``Covariant field equations of the M-theory five-brane,''
  Phys.\ Lett.\  B {\bf 399}, 49 (1997)
  [arXiv:hep-th/9702008].
  
\bibitem{Pasti:1997gx}
  P.~Pasti, D.~P.~Sorokin and M.~Tonin,
  ``Covariant action for a D = 11 five-brane with the chiral field,''
  Phys.\ Lett.\  B {\bf 398}, 41 (1997)
  [arXiv:hep-th/9701037].

\bibitem{oldM5-2}
  I.~A.~Bandos, K.~Lechner, A.~Nurmagambetov, P.~Pasti, D.~P.~Sorokin and M.~Tonin,
  ``Covariant action for the super-five-brane of M-theory,''
  Phys.\ Rev.\ Lett.\  {\bf 78}, 4332 (1997)
  [arXiv:hep-th/9701149].
  M.~Aganagic, J.~Park, C.~Popescu and J.~H.~Schwarz,
  ``World-volume action of the M-theory five-brane,''
  Nucl.\ Phys.\  B {\bf 496}, 191 (1997)
  [arXiv:hep-th/9701166].
  I.~A.~Bandos, K.~Lechner, A.~Nurmagambetov, P.~Pasti, D.~P.~Sorokin and M.~Tonin,
  ``On the equivalence of different formulations of the M theory  five-brane,''
  Phys.\ Lett.\  B {\bf 408}, 135 (1997)
  [arXiv:hep-th/9703127].

\bibitem{M51}
  P.~M.~Ho and Y.~Matsuo,
  ``M5 from M2,''
  JHEP {\bf 0806}, 105 (2008)
  [arXiv:0804.3629 [hep-th]].

\bibitem{M52}
  P.~M.~Ho, Y.~Imamura, Y.~Matsuo and S.~Shiba,
  ``M5-brane in three-form flux and multiple M2-branes,''
  JHEP {\bf 0808}, 014 (2008)
  [arXiv:0805.2898 [hep-th]].
  
\bibitem{BL}
  J.~Bagger and N.~Lambert,
  ``Modeling multiple M2's,
  Phys.\ Rev.\  D {\bf 75}, 045020 (2007)
  [arXiv:hep-th/0611108]; J.~Bagger and N.~Lambert,
  ``Gauge Symmetry and Supersymmetry of Multiple M2-Branes,''
  Phys.\ Rev.\  D {\bf 77}, 065008 (2008)
  [arXiv:0711.0955 [hep-th]].
  J.~Bagger and N.~Lambert,
  ``Comments On Multiple M2-branes,''
  JHEP {\bf 0802}, 105 (2008)
  [arXiv:0712.3738 [hep-th]].

\bibitem{G}
  A.~Gustavsson,
  ``Algebraic structures on parallel M2-branes,''
  Nucl.\ Phys.\  B {\bf 811}, 66 (2009)
  [arXiv:0709.1260 [hep-th]].


\bibitem{Cornalba:2002cu}
  L.~Cornalba, M.~S.~Costa and R.~Schiappa,
  ``D-brane dynamics in constant Ramond-Ramond potentials and  noncommutative
  geometry,''
  Adv.\ Theor.\ Math.\ Phys.\  {\bf 9}, 355 (2005)
  [arXiv:hep-th/0209164].

\bibitem{Ooguri:2003qp}
  H.~Ooguri and C.~Vafa,
  ``The C-deformation of gluino and non-planar diagrams,''
  Adv.\ Theor.\ Math.\ Phys.\  {\bf 7}, 53 (2003)
  [arXiv:hep-th/0302109].

\bibitem{deBoer:2003dn}
  J.~de Boer, P.~A.~Grassi and P.~van Nieuwenhuizen,
  ``Non-commutative superspace from string theory,''
  Phys.\ Lett.\  B {\bf 574}, 98 (2003)
  [arXiv:hep-th/0302078].

  

\bibitem{Nambu}
  Y.~Nambu,
  ``Generalized Hamiltonian dynamics,''
  Phys.\ Rev.\  D {\bf 7}, 2405 (1973);
  F. Bayen, M. Flato, 
  ``Remarks concerning Nambu's generalized mechanics,'' 
  Phys.\ Rev.\ D {\bf 11}, 3049 (1975);
  N. Mukunda, E. Sudarshan,
  ``Relations between Nambu and Hamiltonian mechanics,''
  Phys.\ Rev.\ D {\bf 13}, 2846 (1976); 
  L.~Takhtajan,
  ``On Foundation Of The Generalized Nambu Mechanics (Second Version),''
Commun.\ Math.\ Phys.\  {\bf 160}, 295 (1994)
  [arXiv:hep-th/9301111]. 
  For a review of the Nambu-Poisson bracket, see: 
  I. Vaisman, 
  ``A survey on Nambu-Poisson brackets,''
  Acta. Math. Univ. Comenianae {\bf 2} (1999), 213.
  
\bibitem{Ho:2007vk}
  P.~M.~Ho and Y.~Matsuo,
  ``A toy model of open membrane field theory in constant 3-form flux,''
  Gen.\ Rel.\ Grav.\  {\bf 39}, 913 (2007)
  [arXiv:hep-th/0701130].
  
\bibitem{Furuuchi:2009zx}
  K.~Furuuchi and T.~Takimi,
  ``String solitons in the M5-brane worldvolume action with Nambu-Poisson
  structure and Seiberg-Witten map,''
  JHEP {\bf 0908}, 050 (2009)
  [arXiv:0906.3172 [hep-th]].
  
\bibitem{HoM5}
  P.~M.~Ho,
  ``A Concise Review on M5-brane in Large C-Field Background,''
  arXiv:0912.0445 [hep-th].
  
\bibitem{ChuSmith}
  C.~S.~Chu and D.~J.~Smith,
  ``Towards the Quantum Geometry of the M5-brane in a Constant $C$-Field from
  Multiple Membranes,''
  JHEP {\bf 0904}, 097 (2009)
  [arXiv:0901.1847 [hep-th]].
\bibitem{Huddleston:2010cx}
  J.~Huddleston,
  ``Relations between M-brane and D-brane quantum geometries,''
  arXiv:1006.5375 [hep-th].
  
\bibitem{Pasti:2009xc}
  P.~Pasti, I.~Samsonov, D.~Sorokin and M.~Tonin,
  ``BLG-motivated Lagrangian formulation for the chiral two-form gauge field in
  D=6 and M5-branes,''
  Phys.\ Rev.\  D {\bf 80}, 086008 (2009)
  [arXiv:0907.4596 [hep-th]].

\bibitem{Furuuchi:2010sp}
  K.~Furuuchi,
  ``Non-Linearly Extended Self-Dual Relations From The Nambu-Bracket
  Description Of M5-Brane In A Constant C-Field Background,''
  JHEP {\bf 1003}, 127 (2010)
  [arXiv:1001.2300 [hep-th]].

\bibitem{Chen:2010ny}
  C.~H.~Chen, K.~Furuuchi, P.~M.~Ho and T.~Takimi,
  ``More on the Nambu-Poisson M5-brane Theory: Scaling limit, background
  independence and an all order solution to the Seiberg-Witten map,''
  arXiv:1006.5291 [hep-th].
  
\bibitem{Bekaert:1999dp}
  X.~Bekaert, M.~Henneaux and A.~Sevrin,
  ``Deformations of chiral two-forms in six dimensions,''
  Phys.\ Lett.\  B {\bf 468}, 228 (1999)
  [arXiv:hep-th/9909094].
  

  
\end{thebibliography}
\end{document}